\documentstyle[preprint,aps,epsf]{revtex}

\begin{document}
\draft
\tighten

\preprint{gr-qc/9706063}
\title{SINGULARITIES INSIDE NON-ABELIAN BLACK HOLES}%
\footnotetext{
 Published in {\it Pis'ma Zh.\ Eksp.\ Teor.\ Fiz.}, {\bf 65}, 855 (1997)
 (in Russian).}

\author{D.~V.~Gal'tsov}
\address{Department of Theoretical Physics,
 Moscow State University, \\ 119899 Moscow, Russia \\
 {\tt galtsov@grg.phys.msu.su}}

\author{E.~E.~Donets}
\address{Laboratory of High Energies, JINR, 141980 Dubna, Russia \\
 {\tt donets@sunhe.jinr.ru}}

\author{M.~Yu.~Zotov}
\address{Skobeltsyn Institute of Nuclear Physics, Moscow State University, \\
 119899 Moscow, Russia \\ {\tt zotov@eas.npi.msu.su}}

\maketitle

\begin{abstract}
 Singularities inside the static spherically symmetric black holes in 
 the SU(2) Einstein--Yang--Mills  and Einstein--Yang--Mills--dilaton 
 theories are investigated. Analytical formulas are presented 
 describing oscillatory and power law metric behavior near 
 spacelike singularities in generic solutions.
\end{abstract}
\pacs{04.20.Jb, 97.60.Lf, 11.15.Kc}

 Black holes solutions in the theories including non-Abelian 
 massless vector fields \cite{bhs} are known to exhibit 
 unusual features, which are not shared by the ``classical''
 vacuum or electrovacuum black holes. Besides violation of the naive 
 no-hair and uniqueness theorems, they demonstrate essentially new type 
 of internal structure \cite{dz}. It was shown \cite{dgz} that the space-time 
 inside a generic static black hole in the pure
 Einstein--Yang--Mills (EYM) theory has no Cauchy 
 horizons while the metric oscillates with an infinitely growing amplitude  
 as the singularity is being approached. Although some
 particular solutions still may possess either Schwarzschild (`S') or
 Reissner--Nordstr\"om (`RN') type singularities \cite{dgz}, 
 these configurations constitute only a zero measure set in the solution space. 
 Inside a {\it generic\/} EYM black hole the mass
 function exhibits successive exponential jumps followed by abrupt 
 falls to almost zero values. The amplitude of jumps is  
 exponentially growing up towards  the singularity, while
 the duration of cycles tends to zero.
 The behavior of the system near the singularity is described well
 by a two-dimensional dynamical system \cite{dgz}, which clearly demonstrates 
 an infinitely oscillating regime. These results 
 were confirmed in \cite{blm}, the existence proof for the RN--type 
 solution was announced in \cite{sw}.

 The purpose of the present paper is to discuss the singularity structure
 in generic static spherical black holes in the theories EYM and EYM 
 with dilaton (EYMD) (see \cite{dil} and references therein).
 We assume the `stringy' value of the dilaton coupling constant
 and without loss of generality choose the units and the length scale 
 in which both the Planck mass and the gauge coupling constant are
 equal to unity:
\begin{equation}
                S = \frac{1}{16\pi} \int \left\{ -R+ 2 (\nabla \phi)^2 - 
                    {\rm e}^{-2 \phi} F^2 \right\} \sqrt{-g} d^4x \,,
\label{eq1}
\end{equation}
 where $F$ is the SU(2) field corresponding to the connection
\[
              A_{\mu}^a T_a dx^\mu = (W(r) - 1)(T_\phi d \theta - T_\theta
              \sin \theta d \phi)\, ,
\]  
 ($T_\phi$ and $T_\theta$ being spherical projections of the SU(2) 
 generators). 

 It is convenient to parametrize the space-time interval as
\begin{equation}
        ds^2 = \frac{\Delta \sigma^2}{r^2} dt^2 - \frac{r^2}{\Delta}dr^2
               -r^2 (d \theta^2 + \sin^2 \theta d \phi^2) \,,
\label{eq2}
\end{equation}
 where $\Delta=r^2 - 2r m(r)$ is negative inside the black hole of the
 generic type.

 The equations of motion for $W$, $\Delta$, $\phi$ can be decoupled from 
 the equation for $\sigma$, so that in the general case of the EYMD
 system the equations take the form
\begin{eqnarray}
     && \Delta U'-2\Delta U \phi' = W V/r-{\cal F}W' \,, \label{eq3} \\
     && (\Delta/r)'+\Delta \phi'^2={\cal F}-2 \Delta U^2 {\rm e}^{-2 \phi} \,,
        \label{eq4} \\
     && (\Delta \phi')' + \Delta r \phi'^3 =
        {\cal F} - 2 \Delta(\phi' r + 1) U^2 {\rm e}^{-2 \phi}-1 \,, 
        \label{eq5}
\end{eqnarray}
 where
\[
     U = \frac{W'}{r}, \quad 
        {\cal F} = 1 - \frac{V^2 {\rm e}^{-2 \phi}}{r^2}, \quad V = W^2-1 \,.
\]

\newpage
 The remaining equation for $\sigma$ reads
\begin{equation}
     (\ln \sigma)' =r \left( \phi'^2 + 2 U^2 {\rm e}^{-2 \phi} \right)\,.
\label{eq6}
\end{equation}

 The black hole solutions are generated by the initial data
 at the events horizon $r_h$: $W_h =W(r_h)$ and  $\phi_h=\phi(r_h)$,
 subject to the condition ${\cal F}_h > 0$.
 They can be labeled by the ADM mass $M$ and the dilaton charge $D=
 - \lim (r^2 \phi')$ for $r \rightarrow \infty$.
 The units for these quantities may conveniently be fixed requiring
 $\phi(\infty) = 0$. Then both $W_h$ and  $\phi_h$ are
 entitled to take the discrete values to fit the asymptotic flatness 
 conditions. For all physical EYMD black holes $M>D$. 

 Let us start with examining the singularity structure of the
 exponentially  oscillatory interior space-time \cite{dgz} for generic
 black holes in the pure EYM theory. The corresponding truncated system reads
\begin{eqnarray}
     && \Delta U' + \left(1 - \frac{V^2}{r^2} \right)W' = \frac{WV}{r} \,,
        \label{eq7} \\
     && \left(\frac{\Delta}{r} \right)' + 2 \Delta U^2 = 1 - \frac{V^2}{r^2}
        \,, \label{eq8} \\
     && \frac{d}{dr^2} \ln \sigma = U^2 \,. \label{eq9}
\end{eqnarray}

 It is generically observed during numerical integration from the events
 horizon towards the origin (apart from some discrete horizon data) that
 the oscillations of $\Delta$ take place so that, when oscillation progress,
 the right side of the Eq.\ (\ref{eq7}) becomes small with respect to the 
 terms at the left side. Neglecting it, one obtains the following
 approximate first integral of the system:
\begin{equation}
     Z = \Delta U \sigma/r = {\rm const} \,, \label{eq10}
\end{equation}
 which relates oscillations of the mass function to variation of
 $\sigma$. Numerical experiments also show that while the YM function $W$
 remains almost constant up to $r=0$, its derivative is still non-zero and
 is rapidly changing on some very small intervals of $r$. The function $U$
 exhibits a step-like behavior being  constant with high accuracy during 
 almost all the chosen oscillation cycle (Fig.\ 1) and then 
 jumping to a greater absolute value corresponding to the next
 cycle. It is clear from (\ref{eq9}) that $\sigma$ is exponentially falling
 down with decreasing $r$ while $U \approx const$, 
 whereas in the tiny intervals
 of $U$--jumps $\sigma $ remains almost unchanged.
 So $\sigma$ tends to zero through an infinite sequence of exponential
 falls with increasing powers in the exponentially decreasing intervals. 
 Combining this with (\ref{eq10}) and the above mentioned properties
 of $U$ one can deduce rather detailed description of the metric behavior. 

 Let us denote by $r_k$ the value of radial coordinate where $\Delta$ has
 $k$-th local maximum. Soon after passing this point, the function
 $U$ stabilizes at some value $U_k$ approximately equal to the doubled
 value at the point of local maximum (similarly, $U$ increases by about
 a factor of two when approaching the local maximum, whereas $\Delta$
 is almost stationary). Then, according to (\ref{eq9}), $\sigma$ is equal to
\[
     \sigma(r) = \sigma(r_k) \exp \left[ U_k^2 (r^2 - r_k^2) \right] \,,
\]
 From (\ref{eq10}) one finds that, while $U_k \approx const$,
\begin{equation}
     \Delta(r) = \frac{\Delta(r_k)}{r_k} \; r \; 
                 \exp \left[ U_k^2 (r^2_k - r^2) \right] \,. \label{eq11}
\end{equation}
 This function falls down with decreasing $r$ until it reaches a local 
 minimum at 
\begin{equation}
     R_k = \frac{1}{\sqrt{2}\;|U_k|} \approx
            \frac{\sqrt{|\Delta(r_k)|}}{2|V(r_k)|} r_k \,. \label{eq12}
\end{equation}
 In what follows, in view of the observed fact that in the course of 
 oscillations of $\Delta$ the YM function $W$ changes insignificantly, 
 we will put $V=const$.

 Therefore, the mass function is inflating exponentially while $r$
 decreases from $r_k$ to $R_k$. After passing $R_k$, an exponential in 
 (\ref{eq11}) becomes of the order of unity,  hence $\Delta$ starts to grow
 linearly, and the mass function $m(r)$ stabilizes at the value 
 $M_k = m(R_k)$. Such a behavior holds until the point of local maximum of
 $\Delta/r^2$ is reached; this takes place when $\Delta \approx -V^2$ at
 the point
\begin{equation} 
     r^*_k \approx \frac{V^2}{|\Delta(r_k)|} r_k
                   \exp \left[-(U_k r_k)^2 \right] \,. \label{eq13}
\end{equation} 
 After this a rapid fall of $|\Delta|$ is observed causing a violent rise
 of $|U|$. Then the term $2\Delta U^2$ in the Eq.\ (\ref{eq8}) becomes
 negligible and consequently at this stage
\begin{equation} 
     U \Delta \approx -V^2 U_k \,, \label{eq14}
\end{equation} 
 while $r$ practically stops. This implies that very soon
 $\Delta$ reaches the next local maximum at the point
 $r_{k+1} \approx r_k^*$, while $m(r)$ rapidly falls down to
 $m_{k+1}$. At the point of local maximum of  $\Delta$ 
 one has in the Eq.\ (\ref{eq8}) $|\Delta| \ll V^2$, 
 then in view of the smallness of $r$ we find
\begin{equation} 
     |U(r_k)|  \approx \frac{|V|}{\sqrt{2|\Delta(r_k)|} r_k} \,. \label{eq15}
\end{equation} 

 To obtain the estimates by the order of magnitude we will neglect
 all numerical coefficients elsewhere except for the power indices 
 in exponentials, in particular putting $U(r_k) = U_k$, and omitting
 also (quasi-constant) factors $V$. With this accuracy one obtains from 
 (\ref{eq11})--(\ref{eq15}):
\[
     r_{k+1} = M_k^{-1} \,, \quad r_{k+1}^2 = R_k R_{k+1} \,,\quad
     M_k = \frac{R_k^2}{r_k^3} \exp \left( \frac{r_k^2}{2 R_k^2} \right) \,,
\]
\[
     |\Delta(r_k)| = \left( \frac{R_k}{r_k} \right)^2 \,, \quad
     \frac{r_{k+1}}{r_k} = \frac{r_k^2}{R_k^2} \exp \left[ -\left( 
     \frac{r_k^2}{2 R_k^2} \right) \right] \,.
\]
 Thus, introducing a variable $x_k=(r_k/R_k)^2 \, (\gg 1)$, we can derive 
 the following recurrent formula
\[
     x_{k+1} = x_k^{-3} {\rm e}^{x_k} \,,
\]
 which shows that $x_k$ is an exponentially diverging sequence.
 In terms of $x_k$ one has
\[
     \frac{r_{k+1}}{r_k} = x_k {\rm e}^{-x_k/2} \,,
\]
 this relation can also be understood as a ratio of the neighboring
 oscillation periods since $r_k \gg r_{k+1}$. Values of the function
 $|\Delta|$ at the points $r_k$ rapidly tend to zero:
\[
     |\Delta(r_k)| = x_k^{-1} \,,
\]
 so we deal with an infinite sequence of ``almost'' Cauchy horizons
 as $r \rightarrow 0$. At the same time the values of  $|\Delta|$ at the
 points $R_k$ grow rapidly:
\[
     |\Delta(R_k)| = x_k^{-3/2} {\rm e}^{x_k/2} \,,
\]
 and the values of the mass function grow correspondingly as
\[
     \frac{M_k}{M_{k-1}} = x_k^{-1}{\rm e}^{x_k/2} \,.
\]

 As was shown in \cite{dgz}, in such a regime the system
 (\ref{eq7})--(\ref{eq8}) can be reduced
 to a two-dimensional autonomous dynamical system, one of the fixed
 points of which is an unstable focus with trajectories infinitely 
 spiraling outward.

 While $r$ decreases form $r_k$ to $R_k$, the function $\sigma$ 
 rapidly falls down to the value $\sigma_k = \sigma(R_k)$, which then
 remains unchanged until the point $r_{k+1}$. As the singularity
 is approached, the sequence  $\sigma_k$ decreases according to
\[
     \frac{\sigma_{k+1}}{\sigma_k} = {\rm e}^{-x_k/2} \,.
\]

 Now consider the singularity structure for generic black holes in the
 EYMD theory. In this case,
 starting numerical integration at the horizon, one does not observe 
 huge oscillations of the metric in the interior region. A generic solution
 does not exhibit Cauchy horizons either, so $\Delta$ remains negative
 for all $0 < r < r_h$. For sufficiently 
 small $r$ the right sides of the Eqs.\ (\ref{eq3})--(\ref{eq5})
 become small in comparison with the left hand side terms and one
 gets the following truncated system
\begin{eqnarray}
     && \left( \ln U \right)'-2\phi' = 0 \,, \nonumber \\
     && \left[ \ln (\Delta/r) \right]' = \left[ \ln (\Delta \phi')
        \right]' = -r \phi'^2 \,. \label{eq16}
\end{eqnarray}
 Its integration gives the following five-parameter (i.e. generic) 
 family of solutions  
\begin{eqnarray}
     && W = W_0 + b r^{2(1 - \lambda)} \,, \nonumber \\
     && \Delta = -2 \mu r^{(1 - \lambda^2)} \,, \label{eq17} \\
     && \phi = c + \ln \left( r^{-\lambda} \right) \,, \nonumber
\end{eqnarray}
 with constant  $W_0$, $b$, $c$, $\mu$, $\lambda$.
 The validity of the truncated equations (\ref{eq16}) now can be checked
 by substituting the asymptotic solution (\ref{eq17}) into the full system
 (\ref{eq3})--(\ref{eq5}). For consistency it is sufficient that the following
 inequalities hold:
\[
     \sqrt{2} - 1 < \lambda < 1 \,,
\]
 which is in agreement with the numerical data.

 From (\ref{eq17}) it follows that the mass function diverges as
 $r \rightarrow 0$ according to the power law:
\[
     m(r) = \frac{\mu}{r^{\lambda^2}} \,.
\]
 The corresponding $\sigma$  tends to zero as
\[
     \sigma(r)= \sigma_1 r^{\lambda^2} \,,
\]
 where $\sigma_1 = \text{const}$. 
 Thus, the dilaton completely suppresses the oscillations of the metric
 inside the black hole. 

 Let us discuss now the geometry of space-time near the singularity.
 The regimes described above correspond to an expectation that
 singularity inside black holes of the general type should be spacelike.
 The metric (\ref{eq2}) in the interior region of a black hole corresponds
 to the anisotropic Kantowski--Sachs cosmological solution. 
 One can show that the corresponding shear parameter $\tilde\sigma$ defining
 the shear tensor $\sigma_{ij}=(2, -1, -1)\tilde \sigma/3$, 
 infinitely grows as the singularity is approached.
 For an oscillating EYM solution the values of  $\tilde \sigma$ at the points
 $R_k$ are of the order of
\[
      \tilde \sigma^{\text{max}}_k \sim \frac{M_k^{1/2}}{R_k^{3/2}} \,,
\]
 while the values at the points $r_k$ are 
\[
     \tilde \sigma^{\text{min}}_k \sim \frac{M_{k-1}}{R_{k-1}} \,.
\]
 Both sequences are infinitely growing.
 For the  EYMD solutions the divergence of  $\tilde \sigma$ is power-like.

 Thus, spherical non-Abelian black holes of the generic type
 are conformable with the strong cosmic censorship principle
 (singularity is spacelike). In the case of the pure EYM theory, there is,
 however, an infinite sequence of ``almost'' Cauchy horizons
 in the vicinity of which the mass function starts to grow
 exponentially stabilizing on the value corresponding to
 the next oscillation cycle; the sequence of these  values diverges
 exponentially as the singularity is approached.
 In the theory including a dilaton the mass function monotonously
 (according to the power law) tends to infinity. In the cosmological
 interpretation such a behavior corresponds to an infinite growth of
 anisotropy.

 D.V.G. thanks the Theory Division, CERN for
 hospitality in December 1996 while the work was in progress.
 The research was  supported in part by the RFBR grants 96--02--18899, 18126.

\begin{figure}[htbp]
 \vspace{2cm}
 \centerline{\epsfxsize=15cm \epsfbox{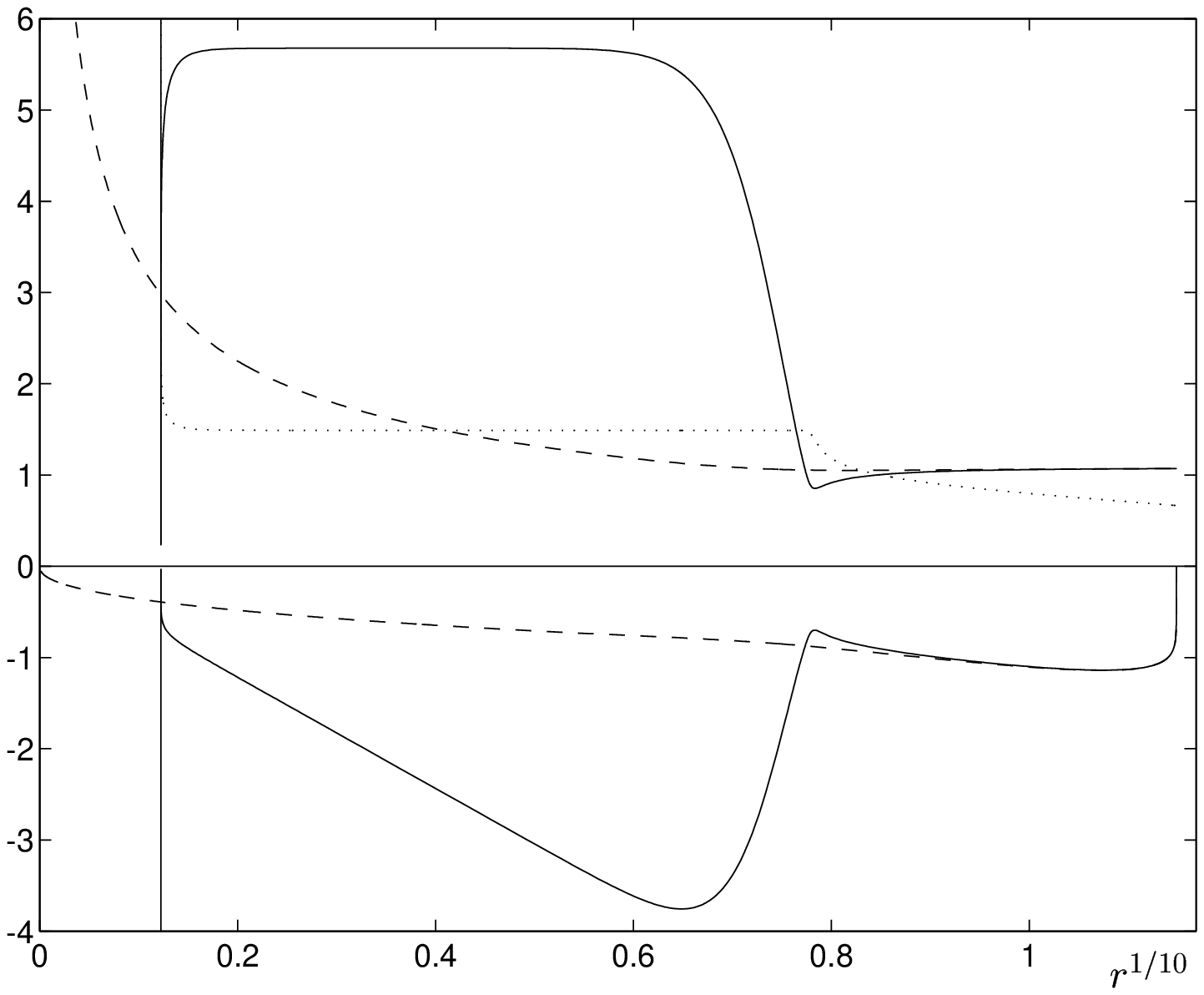}}
 \vspace{5mm}
 \caption{
    The first oscillation cycle for an EYM solution. 
    The functions $\Delta$ for EYM (solid line) and EYMD (dashed)
    are shown in lower half-plane. In upper half-plane~---
    mass functions $m(r)$ (analogously) and the function $U$ for EYM (dotted). 
    All functions are power rescaled with the power index 1/10.
    Here $r_h = 4$; $W_h = -0.283993$ for EYM,
    $W_h = -0.298357$, $\phi_h = 0.05623$ for EYMD
    (asymptotically flat solutions with one node of $W$.)}
\end{figure}

\begin{figure}[htbp]
 \vspace{2cm}
 \centerline{\epsfxsize=15cm  \epsfbox{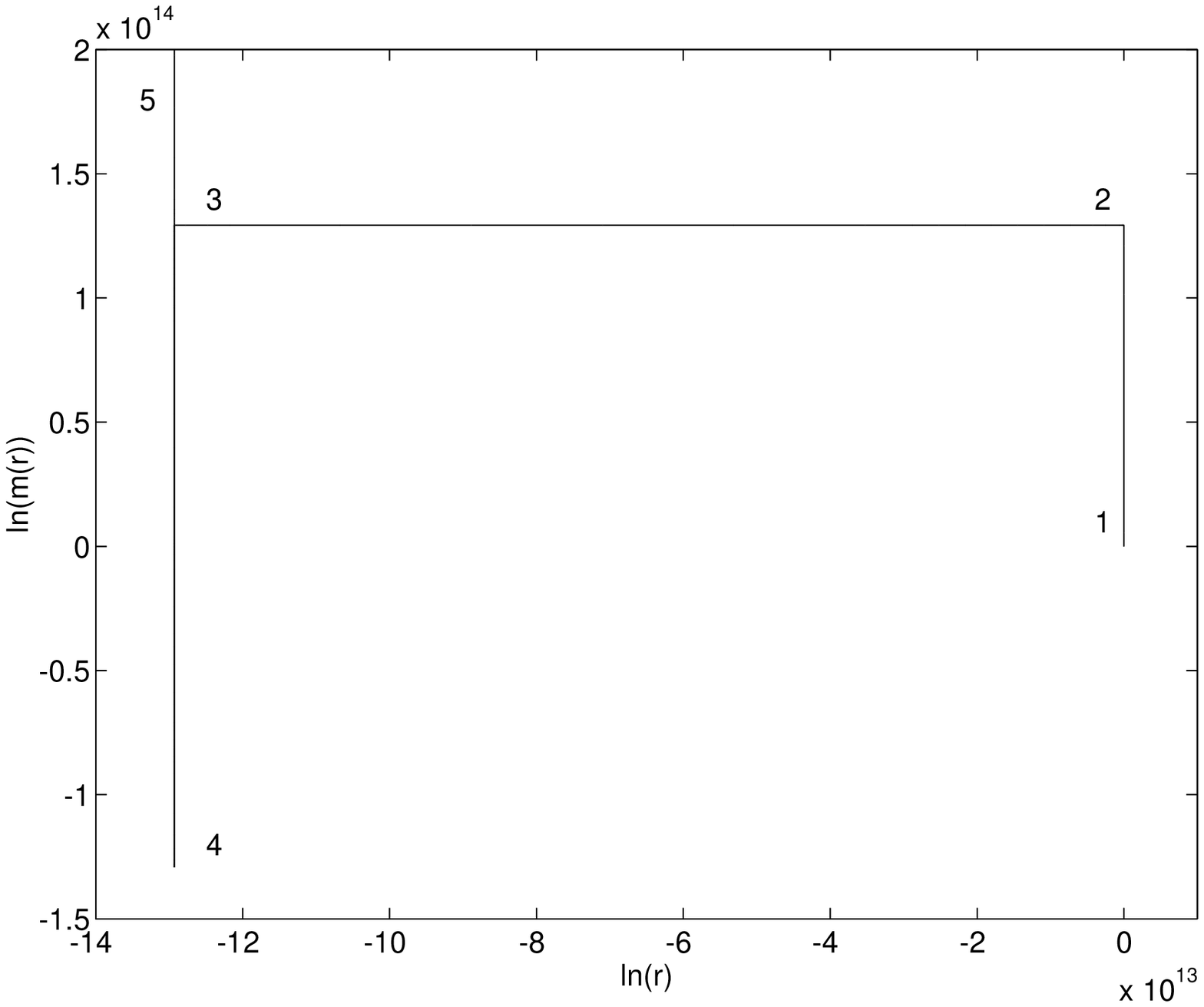}}
 \vspace{5mm}
 \caption{The second oscillation cycle for the EYM mass function.}
\end{figure}


\begin{references}


\bibitem{bhs}
 M.~S.~Volkov, D.~V.~Gal'tsov, Pis'ma Zh.\  Eksp.\  Teor.\  Fiz.,
 {\bf 50}, 312 (1989) [JETP Lett., {\bf 50}, 345 (1990)];
 H.~P.~Kunzle, A.~K.~M.~Masood--ul--Alam, J.\  Math.\  Phys.\ 
 {\bf 31}, 928 (1990);
 P.~Bizon, Phys.\  Rev.\  Lett., {\bf 64}, 2844 (1990).

\bibitem{dz} 
 D.~V.~Gal'tsov, M.~Yu.~Zotov, {\it Interior Solutions
 of the Einstein-Yang-Mills Equations}, in ``Foundations
 of Gravitation and Cosmology'', Abstracts of the International
 School-Seminar, Odessa, September 1995; {\it New Asymptotics for
 the Einstein--Yang--Mills Equations}, in ``Theoretical and Experimental
 Problems of Relativity and Gravitation'', Abstracts of
 the 9th Russian  Gravitational Conference, Novgorod, July 1996.

\bibitem{dgz} 
 E.~E.~Donets, D.~V.~Gal'tsov, and M.~Yu.~Zotov, 
 {\it Internal Structure of Einstein--Yang--Mills Black Holes},
 Preprint DTP-MSU/96-41, gr-qc/9612067.

\bibitem{blm}
 P.~Breitenlohner, G.~Lavrelashvili, and D.~Maison,
 {\it Mass inflation and chaotic behavior inside hairy black holes},
 MPI-PhT/97-20, gr-qc/9703047.

\bibitem{sw}
 J.~A.~Smoller and A.~G.~Wasserman {\it Reissner--Nordstr\"om--like
 solution of the SU(2) Einstein--Yang/Mills equations}, gr-qc/9703062.

\bibitem{dil} 
 D.~V.~Gal'tsov and E.~E.~Donets, Int.\ J.\ Mod.\ Phys.\ {\bf D3}, 755 (1994).

\end{references}
\end{document}